\documentclass[12pt]{revtex4}
\usepackage{epsfig}
\usepackage{graphicx}
\normalsize

\begin{document}

\title{Thermo-kinetic approach of single-particles and clusters involving anomalous diffusion under viscoelastic response}

\author{I. Santamar\'{\i}a-Holek$^{*}$, J. M. Rub\'{\i}\dag , A. Gadomski\ddag }

\address{$^{*}$Facultad de Ciencias, Universidad Nacional Aut\'{o}noma de M\'{e}xico.\\
Circuito exterior de Ciudad Universitaria. 04510, M\'{e}xico D. F.,
M\'{e}xico}
\address{\dag Facultat de F\'{\i}sica, Universitat de Barcelona. \\
Av. Diagonal 647, 08028, Barcelona, Spain}
\address{\ddag Institute of Mathematics and Physics, \\University of Technology and Natural Sciences,
Kaliskiego 7, 85-796 Bydgoszcz, Poland} 


\newpage

\begin{abstract}
\end{abstract}

\maketitle 
\newpage

ABSTRACT
\vspace{2.5cm}

We present a thermo-kinetic description of anomalous diffusion of single-particles and
clusters in a viscoelastic medium in terms of a non-Markovian diffusion
equation involving memory functions. The scaling behaviour of these functions is analyzed
by considering hydrodynamics and cluster-size space random walk arguments.
We explain experimental results on diffusion of Brownian particles in the cytoskeleton, in cluster-cluster aggregation and in a suspension of
micelles.

\vspace{2cm}


\newpage

\section{INTRODUCTION}

Diffusion is an irreversible phenomenon which 
arises in systems of different natures. The understanding of the 
mechanisms intrinsic to this process remains an open question in 
cases where the host liquid displays 
a variety of time and length scales. Such cases include diffusion in slow
relaxation systems \cite{WEITZNATURE06,Vekilov,Oxtoby,wong}, in viscoelastic
media  \cite{lubensky,caspi3,storm,finite} or  in the presence of soft matter aggregations \cite{Vekilov,Oxtoby,adam2000,growing,WEITZ_nuclAndgr,Tanaka}. 
To find out the peculiar behaviour of the mean square displacement (MSD)
of the particles observed in these situations is a theme of vital
importance in areas such as soft condensed matter, biophysics and
material science.

The properties of viscoelastic fluids can be characterized through microrheological experiments in which the MSD of passively diffusing
particles is measured \cite{wong,lubensky,storm}. For example, it has been
shown both experimentally \cite{wong} and theoretically \cite{finite}
that the MSD of a probe particle moving through a medium made up of
a solution of F-actin networks obeys a power law, with an exponent depending on the aspect ratio between the particle radius and the characteristic length of the polymer network. Similar results for the MSD have also been obtained in the case of living cells \cite{caspi3} and then used to explain the viscoelastic properties of the intracellular medium. This explanation is usually given in terms of the single-point generalized Stokes-Einstein relation in the frequency domain \cite{caspi3,weitz82,SteadyState,nonmarkov}. 
The viscoelastic and non-Newtonian properties
are modif\mbox{}ied when the system is subjected to
shear stresses or temperature gradients \cite{david}, and to geometrical
conf\mbox{}inement \cite{Vekilov,david2} in which the finite size of the particles may also play an important role in the dynamics of the system \cite{finite,polonica}.

Matter aggregation phenomena in viscoelastic
environments show up many other interesting features 
both at 
molecular as well as supramolecular or mesoscopic levels of organization
\cite{growing,PNAS,PNAS1}. Although the molecular level is by now
thoroughly accessible through modern experimental techniques
and computer simulations, its mesoscopic counterpart lends 
itself readily
amenable to theoretical modeling, regardless of whether 
the process under study reflects subdiffusive or superdiffusive
dynamic properties \cite{finite,nonmarkov}. As for the former, one
could mention formations of networks and gels
\cite{growing,SAWR1}, viscoelastic phase separation
\cite{Tanaka,Tanaka2} as well as formations of biomembranes and/or soft
biomaterials \cite{adam2000}. As for the latter, one could point
out colloidal and protein systems in which nucleation-growth
phase transformations often lead to a global superdiffusive
behavior, especially when formation of non-Kossel protein crystals
or colloidal crystals comes into play 
\cite{polonica,adam06INPRESS}. In such
systems, f\mbox{}initeness \cite{finite,polonica} and mass flucutations \cite{Ausloos} of the basic constituent (a large Brownian-type
particle, or a cluster of smaller particles) as well as some
detected viscoelastic response of the system upon aggregation are
of prior signif\mbox{}icance and point undoubtedly
to more realistic descriptions.

In this paper, we propose a thermo-kinetic framework based on
mesoscopic nonequilibrium thermodynamics \cite{PNAS} giving a
coarse-grained description of the anomalous transport of particles
and their clusters in
a viscoelastic medium through a non-Markovian diffusion equation, 
engaging thoroughly the slowest dynamic variable termed the reaction coordinate \cite{PNAS1}.
This equation is derived under the assumption of a
non-instantaneous response of the system in which transport
coefficients become memory functions
\cite{zwanzig,trimper,balescu}. These coefficients are introduced into the description
through the expression of the probability current obtained
from the entropy production equation \cite{PNAS,PNAS1}.

The paper is organized as follows. In section \textbf{II} we will
derive the non-Markovian diffusion equation accounting for
the dynamics of the particles in
a viscoelastic medium. In Sec. \textbf{III} we will analyze the 
scaling of the memory function in a polymer solution and in 
Sec. \textbf{IV} we will analyze the scaling of the MSD
in cluster-cluster aggregation. Section \textbf{V} will be devoted to
comparing theory with experiments and Sec. \textbf{VI} to discussion 
and summary of our main results.

\section{The non-Markovian diffusion equation}

Let us consider the Brownian motion of a particle
through a viscoelastic medium such as a polymer melt in which the
macromolecules are distributed as to form a network which 
exerts elastic
forces on the particle \cite{lubensky,caspi3,storm}. In general,
the medium may also involve dynamical processes
associated with changes in the conformation and the growth of its 
costitutents as is the case of the polymerization of
microtubules or the aggregation and
crystallization of lipids and proteins
\cite{caspi3,adam2000,adam06INPRESS,growing}.
For simplicity we will obtain the properties of this
viscoelastic heat bath under an effective medium approximation
\cite{lubensky,finite}.

The dynamics of the particle can be described \cite{PNAS,PNAS1,degroot} 
by means of the
probability density $\rho({\bf r},t)$ depending on its position
${\bf r}$ and time $t$. Since $\rho$ is normalized to
the unity it obeys the continuity equation
\begin{equation}
\frac{\partial }{\partial t}\rho = - \nabla \cdot {\bf J},
\label{continuidadP}
\end{equation}
where $\nabla =\frac{\partial }{\partial {\bf r}}$. The explicit
form of the diffusion current  ${\bf J}$ can be inferred from
the entropy production of the diffusion process.
This quantity can be obtained from the nonequilibrium entropy \cite{PNAS,PNAS1,degroot}
\begin{equation}
s(t)=-k_{B}\int \rho({\bf r},t) \ln
\frac{\rho}{\rho_{le}}d{\bf r}+ s_{le},  \label{p. gibbs}
\end{equation}
where $k_{B}$ is Boltzmann's constant and
$s_{le}$ its value at the reference local equilibrium state characterized by the probability density
\begin{equation}
\rho_{le}({\bf r})=e^{\frac{1}{k_{B}T}[\mu _{le}-\phi_{T}]}.
\label{pleq}
\end{equation}
Here $\mu _{le}$ is the corresponding chemical potential and
$\phi_{T}({\bf r})$ the interaction potential
which contains two contributions: $\phi_{e} $
related to the potential energy of the external forces
and $\phi_{B}({\bf r})$ resulting from the interactions between the
particle and the viscoelastic medium. 

By taking the time derivative of Eq. (\ref{p. gibbs}), using
(\ref{continuidadP}) and integrating by parts we obtain the entropy production
\begin{equation}
\sigma =-\frac{1}{T}\int {\bf J}\cdot \nabla \mu \,d{\bf r},
\label{sigma1part}
\end{equation}
where we have def\mbox{}ined the nonequilibrium chemical
potential: $\mu({\bf r},t)=k_{B}T \ln \rho/\rho_{le}+\mu_{le}$. A
similar type of chemical potential was obtained during a
structural transition of mono-olein bicontinuous cubic phase
induced by inclusion of protein lysozyme solutions, cf. Ref.
\cite{TANAKA3}. In this case, the ratio of nonequilibrium
and equilibrium densities $\rho /{\rho _{le}}$ is
interpreted as a confinement factor which
corresponds to the ratio of the volume of water channels,
$v_w$, and the volume available for the center of lysozyme
molecules $v_a$, namely ${v_w}/{v_a}$, cf. Table II and Eq. (9)
therein.  The structural frustration is caused by a moderate
attraction of the lysozyme molecules mediated by the lipid system
\cite{adam2000}, and can be attributed to a (viscoelastic) phase
separation \cite{Tanaka}.

Following now the scheme of nonequilibrium thermodynamics
\cite{degroot}, we may establish a relationship between the
probability current ${\bf J}$ and  its conjugated force
$\nabla \mu$. Taking into account that in a viscoelastic medium
the response of the system is, in general, not instantaneous
\cite{zwanzig} we obtain
\begin{equation}
{\bf J}({\bf r},t) =- \int_0^t  \zeta(t-t')\rho({\bf r},t')\left(\nabla \phi_T + {
k_{B}T}  \nabla \ln \rho({\bf r},t') \right) dt', \label{zeta}
\end{equation}
where we have used the explicit expression of $\rho_{leq}$ and
introduced the memory function $\zeta (t-t')$. Notice that in the case of non-isotropic systems $\zeta(t-t')$ becomes a tensor.

After substituting Eq. (\ref{zeta}) into (\ref{continuidadP}),  one
obtains the non-Markovian diffusion equation
\begin{equation}  \label{Smol-BP finita}
\frac{\partial \rho({\bf r},t)}{\partial t}= \int^{t}_{0} \zeta (t-t')\nabla \cdot
\left[\rho({\bf r},t') \nabla \phi_{T}\right]dt'+ \int^{t}_{0}D(t-t')
\nabla^2 \rho({\bf r},t') dt'  ,
\end{equation}
where we have defined the memory function $D(t) = k_B T \zeta (t)$. The f\mbox{}irst term on the right-hand side of Eq. (\ref{Smol-BP
finita}) contains the force exerted by the viscoelastic medium on
the particle. In the case of diffusion in an
intracellular medium, there may be elastic forces
arising from the polymer network and the forces due
to the activity of molecular motors
\cite{wong,polonica,luczka}. In the case of diffusion-limited
growth processes  \cite{adam2000}, these forces could also be of an elastic nature or be related to the presence of electrostatic fields \cite{polonica,adam06INPRESS}. The diffusion coefficient obeys the Stokes-Einstein law
\begin{equation}  \label{Dtilde}
D(\omega)={k_{B}T}\beta^{-1}(\omega),
\end{equation}
where $\beta(\omega)=6 \pi a \eta(\omega)$ is the
friction coefficient containing the radius $a$ of the particle
and the frequency-dependent viscosity $ \eta(\omega)$,
\cite{nonmarkov,WEITZ_JofRheol01}. The
memory function $\zeta(t)$ and $\beta(\omega)$ are related through the
inverse Laplace transform: 
$\zeta(t)=\pounds^{-1}\left[\beta^{-1}(\omega) \right]$.  Equation (\ref{Smol-BP finita}) is similar to the
ones previously derived in the context of the continuous-time random
walk formalism and of projector operator techniques
\cite{zwanzig,balescu}.

Explicit expressions for the memory functions $\zeta(t)$ and
$D(t)$ can be obtained from hydrodynamics. It has been shown
\cite{landau FM,RUBI-AGUSTIN} that the frequency-dependent
corrections to the friction coeff\mbox{}icient are given by a power
expansion of the inverse penetration length: $\alpha_0 =
({\omega }/{\nu_0})^{1/2}$, with $\nu_0$ the kinematic viscosity
of the heat bath \cite{landau FM}. In the case when the heat bath
is a viscoelastic fluid, the inverse penetration length
incorporates the frequency dependent kinematic viscosity
$\nu(\omega)$: $\alpha(\omega) = [{\omega }/{\nu(\omega)}]^{1/2}$,
leading to the expression \cite{SteadyState,landau FM,RUBI-AGUSTIN}
\begin{equation}
\beta(\omega) \simeq \beta _{0}\left[ 1+a\alpha(\omega)\right].
\label{beta fax}
\end{equation}

This length behaves as: 
$\alpha(\omega)\sim(a
\Delta_{\omega})^{-1}(\tau_D\omega)^{\left(\frac{1-\delta}{2}\right)}$,
where $\tau_D$ is a characteristic relaxation time. We have assumed that, within a given range of frequencies,
the kinematic viscosity of the viscoelastic medium obeys the
scaling: $\nu(\omega) \simeq \nu_0 \left(\tau_D
\omega\right)^{\delta}$ \cite{SteadyState}. The quantities
$\delta$ and $\Delta_{\omega}$ may, in general, depend on the free
space around the particle \cite{danch} and on the
temperature \cite{SteadyState,TODD}. The scaling factor $\Delta_{\omega}$ has its origin in the fact that the relaxation time of
the system may depend on the concentration of, for instance, polymers or micelles making up the viscoeastic medium \cite{SteadyState,TODD,ferry}.

The exponent $\delta$ also admits different interpretations. An
important one of these is the one related to the Brownian motion of clusters or
small crystals which are still in their late growing stage, e.g.
when forming semi-orderly amorphous mesostructures 
\cite{danch}. In this
case, the exponent may reflect the specif\mbox{}ic dynamics obeyed by the
radius of these objects as has been analyzed in Ref.
 \cite{growing}.

\section{Scaling of the memory function}

The scaling of the memory function and, consequently, the
microrheological properties of the viscoelastic medium can be
inferred from the MSD of a test particle. The MSD can be obtained
experimentally by using diffusing-wave spectrometry
techniques or video-based methods \cite{wong,caspi3,weitz82}. It
has been found that, in a certain range of frequencies, the
complex shear modulus obeys the scaling behavior: $G_{msd}^{\prime
\prime}(\omega) \sim \omega^{\gamma}$ with $\gamma < 1$,
\cite{wong,weitz82}.

These experimental results can be described by using Eq.
(\ref{Smol-BP finita}) in the case of $\phi_T=0$. For frequencies
lying in the range $\tau_D^{-1}<\omega\ll \beta_0$, the leading
term of the Eq. (\ref{beta fax}) yields
\begin{equation}
\beta^{-1}(\omega) \simeq \beta^{-1}_0 \left(\tau_D
\omega\right)^{-\left( \frac{1-\delta}{2}\right)}.
\label{viscoelastic-mobilityfrequency}
\end{equation}
The inverse Laplace transform of
(\ref{viscoelastic-mobilityfrequency}) can be performed by using
the Tauberian theorem \cite{balescu}. One then obtains the memory
function
\begin{equation}
\zeta(t) \simeq \frac{1}{\beta_0 \tau_D} \frac{1}{\Gamma(\frac{%
1-\delta}{2})} \left(\frac{t}{\tau_D}\right)^{-\left(\frac{1+\delta}{2}%
\right)},  \label{viscoelastic-mobilitytime}
\end{equation}
where $\Gamma(b)$ is Euler's gamma function. After
substituting (\ref{viscoelastic-mobilitytime}) into (\ref{Smol-BP
finita}) with $\phi_T=0$ and using the inverse Laplace transform of Eq. (\ref{Dtilde}), one
arrives at the non-Markovian diffusion equation
\begin{equation}  \label{MNdiffusion-linear-short-times}
\frac{\partial \rho}{\partial t}= \frac{D_0%
}{\tau_D \Gamma(\frac{1-\delta}{2})} \int^{t}_{0} \left(%
\frac{t-t'}{\tau_D}\right)^{-\left(\frac{1+\delta}{2}\right)} \nabla^2 \rho(%
{\bf r},t') dt' ,
\end{equation}
which is similar to the one previously derived in Ref.
\cite{balescu}.

It is important to realize that equation
(\ref{MNdiffusion-linear-short-times}) can be recast in the 
form of a 
fractional diffusion equation \cite{FFPE} by Laplace
transforming it
and multiplying the result by $(s\tau_D)^{-1}$. Then, taking the
time derivative of the inverse Laplace
transform of the resulting expression, one obtains
\begin{equation}  \label{FFPE}
\frac{\partial \rho}{\partial t}({\bf r},t)=\, _0D_t^{1-\hat{\alpha}}D_0 \nabla^2 \rho(%
{\bf r},t)  ,
\end{equation}
in which the exponent is $\hat{\alpha}=(3- \delta)/2$, with $1<\delta<3$ and the  integro-differential operator 
\begin{equation}  \label{fraccionaria}
_0D_t^{1-\hat{\alpha}}A(t')= \Gamma^{-1}(\hat{\alpha})\frac{\partial}{\partial t} \int^{t}_{0}\left(%
\frac{\tau_D}{t-t'}\right)^{\left(1-\hat{\alpha}\right)}A(t')dt'
\end{equation}
entering Eq. (\ref{FFPE}) is the zero order Riemann-Liouville operator \cite{FFPE}.
This feature demonstrates that our method based on a scaling of the memory
function and a description using fractional derivatives leads to an identical
kinetic equation.

The value of the exponent $\delta$ characterizing the properties
of the viscoelastic medium can in general be obtained by
analyzing the hydrodynamic effects \cite{finite}.
For simplicity, we will show how this exponent may be inferred
from the evolution in time of the MSD
\begin{equation}
\langle r^2\rangle(t) = \int r^2 \rho({\bf r},t) d{\bf r}.
\label{rsquare}
\end{equation}
The evolution equation for $\langle r^2\rangle(t)$ follows by taking
the time derivative of Eq. (\ref{rsquare}) and then using 
Eq. (\ref{MNdiffusion-linear-short-times}). After integrating by
parts, one obtains
\begin{equation}
\frac{d\langle r^2\rangle(t)}{dt} = \frac{D_0}{%
(1-\delta) \Gamma(\frac{1-\delta}{2})} \left(\frac{t}{\tau_D}\right)^{\frac{%
1-\delta}{2}},  \label{derivativersquare}
\end{equation}
whose solution is
\begin{equation}
\langle r^2\rangle(t) \simeq t^{\frac{3-\delta}{2}}.  \label{MSD}
\end{equation}

Equation (\ref{MSD}) can now be compared with the
experimental result $\langle r^2\rangle_{exp} \simeq
t^{\frac{3}{4}}$, obtained for a Brownian
particle in the intracellular medium \cite{caspi3}, yielding $\delta = 3/2$. In
view of Eq. (\ref{viscoelastic-mobilityfrequency}), the mobility
is given by $\beta(\omega) \simeq \omega^{-%
\frac{1}{4}}$. Since $\beta(\omega) \propto \eta_{eff}(\omega)$ with
$\eta_{eff}(\omega)$ the effective viscosity of the medium and $%
G_{MSD}^{\prime \prime }(\omega)\simeq \omega \eta_{eff}$, it
follows that our
result is in good agreement with the experimental result $%
G_{MSD}^{\prime \prime} \sim \omega^{\frac{3}{4}}$, reported in
Ref. \cite{weitz82}.

\section{Scaling of the MSD in cluster-cluster aggregation}

Cluster-cluster aggregation in a viscoelastic
medium at high temperatures can be viewed as a
$d$-dimensional random walk (RW) in the space of cluster sizes
$x$, \cite{growing}. Below, we present some arguments which 
lead to the recovery of the experimentally justified result given by
Eq. (\ref{MSD}) also at the level of clusters' RW.

A coupling of late-time growing and mechanical
relaxation conditions has been proposed in Ref. \cite{growing} in order
to properly describe the clusters' RW in the fluctuating viscoelastic
medium. The role of MSD for such RW is
played by the average radius of the clusters but squared,
${R_{av}}^2 (t)$, which scales with the time $t$ as \cite{adam2000,growing,SAWR1}
\begin{equation}
{R_{av}}^2 (t) \sim t^{2/(d+1)}.
\label{ad1}
\end{equation}

According to Ref. \cite{growing}, and based on the above scaling
relation, for a planar geometry the value of the exponent of the MSD is $\nu _{MSD} =2/3$,
which is the harmonic mean of two other exponents obtained for $d=1$
and $d=3$, namely $\nu_{MSD}=1$
and $\nu _{MSD} =1/2$, respectively. This is due to some physical
interpretation that the harmonic mean well describes an optimal
(excluded-volume involving \cite{danch}) kinetic pathway for a clustering thereby
introduced, which ultimately favors $d=2$. This interpretation is in
accordance with soft-glass matter-organization conditions which
favor characteristic exponents less than $3/4$, see Ref.
\cite{WEITZ_JofRheol01} and references therein. If we,
however, take an arithmetic instead of the harmonic mean, and apply it
to both extreme exponents, $\nu _{MSD} =1$ and $\nu _{MSD} =1/2$,
which looks like a customary f\mbox{}itting procedure applied by
experimenters, cf. Ref. \cite{WEITZNATURE06}, we f\mbox{}inally arrive at the arithmetic-mean
exponent: ${\nu_{MSD}}^{AM} = 3/4$, which is characteristic of the
viscoelastic matter relaxation (under late-stage growing
conditions) \cite{adam2000,growing}.  This relation
supports the general view offered by the present study.

The scaling formula (\ref{ad1}) can be used to obtain
the particle MSD of specific concentrated colloidal systems.
For example, in the case of the sol-gel continuous phase
transition reported in Ref. \cite{Schurt}
the exponent of the initial and final phases can be acceptably
well-reproduced by taking into account the pivotal role of
clusters' surface $x^{(d-1)/d}$,
\cite{WEITZ_nuclAndgr,growing,SAWR1}. In our model, the
characteristic exponent $\nu _{char}$ of these phases correspond
to $\nu _{char} = 1$ for $d=1$ (no surface) and $\nu _{char} =
0.66$ for $d=2$ (a planar geometry) which is very close to the
measured $0.7$, see Figure \textbf{4} of Ref. \cite{Schurt}.
Thereby, experiment and theory show almost the same
exponents values under which the same dynamic behavior, i.e. a
passage from a diffusive (sol) to some subdiffusive (gel) state
can be seen \cite{growing}. This is also confirmed by relaxation
measurements, cf. Figure \textbf{3} of the review \cite{Schurt} and
our considerations below.

The MSD $R_{av}^2$  introduced in Eq.(\ref{ad1}) can be further related
with the creep compliance $\kappa(t)$, def\mbox{}ined by the average
strain to be assigned to a probe particle immersed in the viscoelastic medium (e.g., the cytoskeleton), divided by the corresponding shear stress
\cite{WEITZ_JofRheol01,ferry}
\begin{equation}
R_{av}^2 (t) \propto \kappa (t),
\label{ad2}
\end{equation}
In the microrheological high-temperature environment
considered in Ref. \cite{growing}, we have made an attempt to
def\mbox{}ine in a phenomenological way the inverse of $\kappa$, namely
$\kappa^{-1}$, as the internal matrix stress $\sigma_M (t)$
accumulated within the inter-cluster spaces of the matrix, assuming
that in the late-stage growing conditions the average strain can
be set constant. Taking this relation to be valid,
the creep compliance for our high $T$ clustering viscoelastic system
satisf\mbox{}ies $\kappa(t) \simeq {\sigma _M}^{-2} (t)$, which after
taking into account Eqs. (\ref{ad1}) and (\ref{ad2}) leads to
\begin{equation}
{\sigma _M} (t) \propto  t^{-1/(d+1)},
\end{equation}
i.e. a slightly modified result than while entirely based on the before
applied phenomenology \cite{growing}. From this relation, it follows that the scaling of the internal
matrix stress will, in general, be a function of the relevant
degree of freedom, $d$ -the space dimension, in the same way as
the MSD.

\section{Comparison with experiments}

\subsection{The cytoskeleton response: fast-slow crossover of the complex shear modulus}

Recent experiments measuring the viscoelastic response of
the cytoskeleton through its complex shear modulus $G''(\omega)$
have shown a crossover from the scaling behaviour $G''(\omega)\sim
\omega^{3/4}$ for large values of the frequency to an almost
constant value for low frequencies \cite{WEITZNATURE06}. These
experiments were performed by attaching a magnetic bead to the
cytoskeleton network and then applying a periodic
magnetic field that exerts a torque on the particle \cite{WEITZNATURE06}.

This crossover can be explained by means of the theory presented in section \textbf{II} if one assumes that the
total force exerted on
the attached probe particle can be modelled by means of the
harmonic force ${\bf F}({\bf r})=-\omega_0^2\bf{r}$ with
$\omega_0$ a characteristic frequency, and
that in Eq. (\ref{viscoelastic-mobilitytime})
$\delta=3/2$.

From our description one may show that the complex
shear modulus is related with $\langle \hat{{\bf r}}^2 \rangle(\omega)$, through the
well-known expression \cite{WEITZ_JofRheol01}
\begin{equation}
G''(\omega)=\frac{k_B T}{\pi a}\frac{1}{\omega \, \langle \hat{r}^2 \rangle},
\label{G-MSD}
\end{equation}
where we have used the relation
$\eta(\omega)=\eta_0 \hat{\zeta}^{-1}(\omega)$ obtained after taking
the Laplace transform of Eq. (\ref{Smol-BP finita}) and using
(\ref{Dtilde}). From the resulting equation
we can obtain the MSD
\begin{equation}
\langle \hat{r}^2 \rangle(\omega)=\frac{k_B T}{\pi a \eta_0}%
\frac{1}{\omega}%
\left(2\omega_0^2 \beta_0^{-1}+\tau_D^{-1/4}\omega^{3/4}\right)^{-1}.
\label{MSDfrequency}
\end{equation}
Substituting now this expression into Eq. (\ref{G-MSD}), one finds
the complex shear modulus of a probe particle
attached to the network
\begin{equation}
G''(\omega)=\eta_0\left(2\omega_0^2 \beta_0^{-1}+
\tau_D^{-1/4}\omega^{3/4}\right).
\label{G-part-adherida}
\end{equation}

As shown in Figure \textbf{1}, our result (\ref{G-part-adherida})
is in good agreement with the experiments reported in Ref.
\cite{WEITZNATURE06}. The linear model introduced in this subsection
proves to be adequate in order to reproduce experiments performed at constant temperature.
The effect of temperature can be incorporated into the
description by taking into account entropic effects in the local
equilibrium distribution function (\ref{pleq}) \cite{david,david2} and through
the dependence on $T$ of the relaxation time
$\tau_D$, \cite{SteadyState,TODD}.

\subsection{Brownian motion in a solution of giant micelles}

We will now proceed to analyze the behavior of the MSD of
a particle undergoing Brownian motion in a semidilute solution
of worm-like micelles \cite{bellour}. The host fluid will be
modelled as a structured viscoelastic medium acting on the
particle with the non-linear force
\begin{equation}
F^{el}(x) = -\nabla \phi = F_0 \cos(\lambda^{-1} x),
\label{potencial}
\end{equation}
where $F_0$ is the amplitude of the force, considering the one-dimensional case for simplicity. $\lambda$ can be
interpreted as the characteritic distance
between the particle and the cage formed by the micelles. 

In order to describe the experimental results for the MSD reported
in \cite{bellour}, we will approximate $\rho({\bf r},t+z)$
through: $\rho({\bf r},t+z)\simeq\rho({\bf
r},t)+(\partial{\rho}/\partial t)(t)z+O(z^{2})$. 
This approximation takes into account that for the
subdiffusive processes we are considering here, the relaxation of 
the system is not exponential and thus no clear separation of times 
scales can be assumed. As a
consequence, at larger times the dynamics of the system remains 
non-Markovian, but with time-dependent coefficients \cite{balescu}. Thus, 
keeping the lowest order term of the expansion and substituting 
the resulting expression into (\ref{Smol-BP finita}), one obtains 
\begin{equation}  \label{Smol-markovianized}
\frac{\partial }{\partial t}\rho(x,t)= {\tilde{D}}(t)
\frac{\partial^{2}}{\partial x^{2}} \rho(x,t)+ \tilde{\zeta}(t)
F_0 \frac{\partial}{\partial x} \left[\, \rho(x,t) \cos\left(
\lambda^{-1} x \right)\right],
\end{equation}
where we have introduced the time-dependent coeff\mbox{}icients
\begin{equation}  \label{beta(t)}
\tilde{\zeta}(t) = \int^t_0 \zeta(z)dz \,\,\,\,\,\,{and}
\,\,\,\,\,\,{\tilde{D}}(t)= {k_B T} \tilde{\zeta}(t),
\end{equation}
where the memory function $\zeta(t)$ is given by Eq.
(\ref{viscoelastic-mobilitytime}).
From Eq. (\ref{Smol-markovianized}) we obtain the
evolution equation for the MSD
\begin{equation}
\frac{d}{d\tau}\langle x^2 \rangle= 2 D_0 - 2 \frac{F_0 D_0}{k_B
T} \langle x \cos\left(\lambda^{-1} x \right) \rangle,
\label{evolMSD-active-Final}
\end{equation}
where we have introduced the new variable $\tau(t)=\int 
\tilde{\zeta}(t)dt$. This equation implies that
the effects due to the non-linear elastic force on the behaviour
of the MSD appear through the average work $\langle x F^{el}
\rangle$ which this force exerts over the particle \cite{david2}.
Since the solution of Eq. (\ref{Smol-markovianized}) with $F^{el}$
given by (\ref{potencial}) cannot be obtained analytically, 
the average $\langle x \cos\left(x/{\lambda}\right)
\rangle$ cannot be explicitly computed. However, we will 
assume small deviations with respect to the average value of the position
of the particle, and then use the following expansion around the mean 
value 
$\langle x\rangle(\tau)$: $\langle
f\left(x\right)\rangle \simeq f\left(\langle
x\rangle\right)+O\left(\langle x^2-\langle
x\rangle^2\rangle\right) $. The resulting equation is
\begin{equation}
\frac{d}{d\tau}\langle x^2 \rangle(\tau)= 2 D_0 - 2 \frac{ F_0
D_0}{k_B T} \langle x\rangle(\tau) \cos\left[\lambda^{-1} \langle
x\rangle(\tau)\right] , \label{evolMSD-active-PROMEDIO}
\end{equation}
which must be solved together with the approximated evolution 
equation for $\langle x\rangle(\tau)$
\begin{equation}
\frac{d}{d\tau}\langle x \rangle(\tau)=  - \frac{F_0 D_0}{k_B T}
\cos\left[\lambda^{-1} \langle x\rangle(\tau)\right] .
\label{evol-PROMEDIO}
\end{equation}

Integrating Eqs. (\ref{evolMSD-active-PROMEDIO}) and
(\ref{evol-PROMEDIO}) one can obtain the following expression
for the MSD
\begin{equation}
\langle x^2 \rangle(t) \simeq \langle x^2 \rangle(t_0)+ 2 D_0
\tau(t) + 4 \lambda^2 \tanh\left[\frac{F_0 D_0}{2 \lambda k_B
T}\tau(t)\right]^2, \label{MSD-2}
\end{equation}
where $t_0$ is a cut-off time,
$\tau(t)=\frac{4}{(3-\delta)(\delta-1)}\Gamma^{-1}\left[(\delta-1)/2\right]
\tau_D^{{(\delta-1)}/{2}}t^{{(3-\delta)}/{2}}$ and, for convenience, 
we have taken the first
term of the expansion of $\arctan(x)^2 \simeq x^2$ with $x=\tanh[\frac{F_0 D_0}{2 \lambda k_B
T}\tau(t)]$. The behaviour of
the MSD (\ref{MSD-2}) as a function of time (solid line) is shown
in Figure \textbf{2}, and compared with experimental data (circles)
taken from Ref. \cite{bellour}. We have used the following values
for the parameters: $D_0 \simeq 3 \cdot 10^{-5}\mu m^2s^{-1}$,
$k_B T \simeq 4.2\cdot 10^{-6}gr \mu m^2s^{-2}$,
$F_0\simeq 0.5gr \mu m\,s^{-2}$, $\tau_D \simeq 0.1 s$ and $\lambda \simeq 8\cdot 10^{-3} \mu m$. From
Figure \textbf{2} we may conclude that the expression for the MSD 
we have obtained agrees with microrheological experiments \cite{bellour}
within the time interval represented. Notice that the plateau 
is a signature of the existence of
cage effects, which in our model become manifest at the 
maximum value
of the elastic force $F^{el}$. The exponent 
characterizing the memory function of the solution of giant micelles 
is $\delta = 9/5$, which differs from that of our previous example.

\section{DISCUSSION}

In this article, we have proposed a thermo-kinetic approach to
analyze anomalous diffusion of particles and clusters in systems having
a viscoelastic response. A non-Markovian diffusion equation has been
derived from the entropy production at the mesoscale related to the diffusion process.

Using hydrodynamic arguments, we have obtained an expression
for the memory function which obeys a power law in frequencies and leads
to the scaling behaviour of the mean square displacement of the Brownian particle. 

Cluster-cluster aggregation in the viscoelastic
medium considered as a
$d$-dimensional random walk in the space of cluster sizes
$x$ has also been analyzed \cite{adam2000,adam06INPRESS}. The role of MSD for such a RW, 
played by the average radius of the clusters but squared, has been
used to argue that the time dependence of the
matrix stress and the creep compliance of the colloid-type system could scale with 
a law depending on the relevant degree of freedom, $d$ -the space dimension of 
the Brownian particle. Moreover, we have shown that the characteristic exponent $3/4$ of the MSD can be obtained by applying an arithmetic mean to both extreme exponents ($d=1$ and $d=3$). This procedure coincides 
well with the one followed when obtaining the exponents from experimental 
data \cite{WEITZNATURE06,WEITZ_JofRheol01}.

As particular cases of our general formalism, we have
studied the dynamics of a particle attached to the cytoskeleton
and of a Brownian particle moving through a semidilute solution of
giant worm-like micelles. In the first case, we have explained the
crossover of the complex shear modulus of the cytoskeleton
observed in experiments in terms of an harmonic-force model. In the second 
case, we have analyzed the dynamics of
a Brownian particle moving through a semidilute solution of giant
worm-like micelles by proposing a nonlinear force model that takes 
cage effects into account. The good agreement between experiment and 
theory, shown through Figures \textbf{1} and \textbf{2}, 
allows one to conclude that the
theoretical approach we have proposed in this article can
explain the behaviour observed in microrheological experiments
\cite{WEITZNATURE06,weitz82,WEITZ_JofRheol01,bellour}. 

Our study suggests for the possibility of providing
justifications of experimental results on diffusion in a viscoelastic 
medium in terms of  a Smoluchowski description \cite{finite,adam06INPRESS}. The theory proposed 
may constitute a useful
tool ready for  examining the dynamics of the
soft-matter systems at a mesoscale.
\\

\begin{large}
\textbf{Acknowledgments}
\end{large}

This work was supported by UNAM-DGAPA under the grant IN-108006
(ISH), and by DGICYT of the Spanish Government under Grant No.
FIS2005-01299. (Dis)ordered aggregation and Smoluchowski type
dynamics aspects concerned with the present study
are subject to a MNiI grant 2P03B
03225(2003-2006) regulations and AG's participation
therein. A short-visit ESF/StochDyn grant is acknowledged too. Thanks go to Prof. R. Rodr\'iguez, Dr. Adam Danch and Dr. Jacek  Si\'odmiak 
for discussions and careful reading of this manuscript.
\\

\newpage

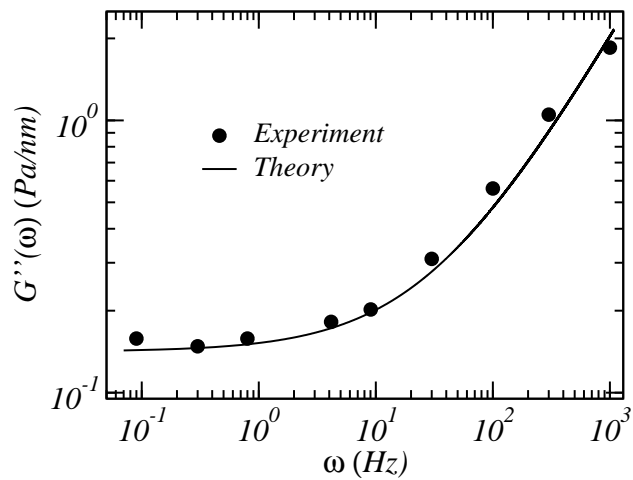
Figure 1.

Complex shear modulus of the cytoskeleton as a function of frequency for the case of a magnetic bead attached to the cytoskeleton network. An oscillatory magnetic field was applied in order to measure the response of the material. Circles represent experimental data taken from Ref. \cite{WEITZNATURE06} whereas the solid line a fit of the data obtained from Eq. (\ref{G-part-adherida}), for $a=2.25\cdot 10^{-6} m$,
$\beta \simeq 1\cdot 10^{6}s^{-1}$, $\eta_0 \simeq 1\cdot 10^{-3}kg m^{-1}s^{-1}$, $\omega_0 \simeq 1.77\cdot 10^{-4}s^{-1}$ and
$\tau_D \simeq 4\cdot 10^{-5}s$.

\vspace{3cm}

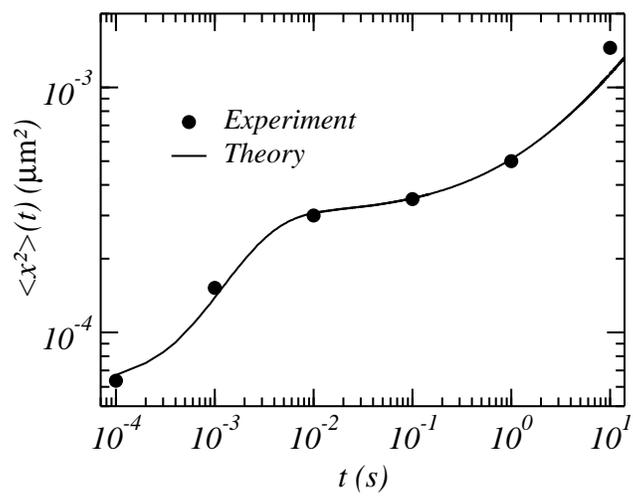
Figure 2.

The MSD of a polystyrene particle in a semidilute
solution of worm-like micelles as a function of time(\ref{potencial}). The black circles are
experimental data obtained by means of diffusing wave spectroscopy techniques, and taken from Ref. \cite{bellour}. The solid
line was obtained with Eq. (\ref{MSD-2}).  The values of the
parameters are given in the text. 

\newpage

\begin{figure}[tbp]
{}
\par
\centering
\mbox{\resizebox*{8.25cm}{!}{\includegraphics{Fig1-kernel.eps}} }
\par
{\footnotesize {\ Figure 1. } \vspace{.8cm} }
\end{figure}

\begin{figure}[tbp]
{}
\par
\centering
\mbox{\resizebox*{8.25cm}{!}{\includegraphics{Fig2.eps}} }
\par
{\footnotesize {\ Figure 2} \vspace{.8cm} }
\end{figure}

\end{document}